\documentclass[11pt]{article}
\usepackage{cospar}
\usepackage[sectionbib]{natbib}
\usepackage{graphicx}
\usepackage[figuresright]{rotating}

\title{PROSPECTS FOR HIGH ENERGY STUDIES OF PULSARS WITH THE AGILE GAMMA-RAY TELESCOPE}

\author{A. Pellizzoni\address{Istituto di Astrofisica Spaziale e Fisica Cosmica (IASF/CNR),
via Bassini, 15, I-20133, Milano, Italy},
        A. Chen$^{1,}$\address{Consorzio Interuniversitario di Fisica Spaziale (CIFS), Torino,
 Italy}, M. Conti$^{1}$, A. Giuliani$^{1, }$$^{2}$, S. Mereghetti$^{1}$, M. Tavani$^{1}$, S. Vercellone$^{1}$
        } 

\begin{document}

\maketitle

\begin{abstract}
AGILE is a small gamma-ray astronomy satellite, with good
spatial resolution, excellent timing capabilities and an unprecedented large field of
view ($\sim$1/5 of the sky). It will be the next mission dedicated to high energy
astrophysics in the range 30 MeV-50 GeV, and will be launched in 2005.
Pulsars are a major topic of investigation of AGILE and, besides studying the small sample
of known objects, AGILE will offer
the first possibility of detecting several young and energetic radio pulsars that have been
discovered since the end of the CGRO mission.
We provide an estimate of the expected number of detectable gamma-ray pulsars and 
present AGILE capabilities for timing analysis
with small counting statistics, based on the analysis of data from simulations, from the 
EGRET archive, and from radio pulsar catalogs.
\end{abstract}

\section*{THE AGILE MISSION}
AGILE is a small scientific mission of the Italian Space Agency dedicated to high-energy
astrophysics (Tavani et al. 2001). It will be the only mission entirely devoted to gamma-ray observations
in the 30 MeV-50 GeV energy band, with simultaneous X-ray imaging also in the 10-40 keV band,
during the period 2005-2007. Despite its small dimensions ($\sim$0.3 m$^3$) and weight ($\sim$100 kg) 
compared to
other future instruments such as GLAST, the AGILE performance will be as good as, or better than,
that of bigger past instruments such as EGRET, thanks to the new silicon detector technology
employed for the AGILE instruments.

High-energy photons are converted into e$^{+}$/e$^{-}$ pairs in a tracker made of
12 Silicon-Tungsten planes, which 
allows us to efficiently collect photons with an effective area of $\sim$500 
cm$^{2}$ and to perform photon direction reconstruction with good angular resolution 
($\sim$0.5$^{\circ}$, $E$$\sim$1 GeV) (Figure 1) on a very large field  of view, covering about 1/5
of the sky in a single pointing (Figure 3). A mini-calorimeter made of Cesium-Iodide bars
supports the tracker for photon energy reconstruction and particle background rejection, 
together with a set of anticoincidence scintillator panels surrounding 5 sides of the instrument.
An additional silicon plane and a coded mask, positioned on top of the tracker, constitute 
SuperAGILE, an X-ray imager with angular resolution of $\sim$1-3 arcmin and sensitivity of $\sim$5 mCrab 
in the band 10-40 keV.
AGILE is characterized by the smallest deadtime ever obtained for gamma-ray detection 
($<$200 $\mu$s) and absolute time tagging with uncertainty near $\sim$1 $\mu$s.


\begin{figure}
\includegraphics[width=95mm,height=100mm]{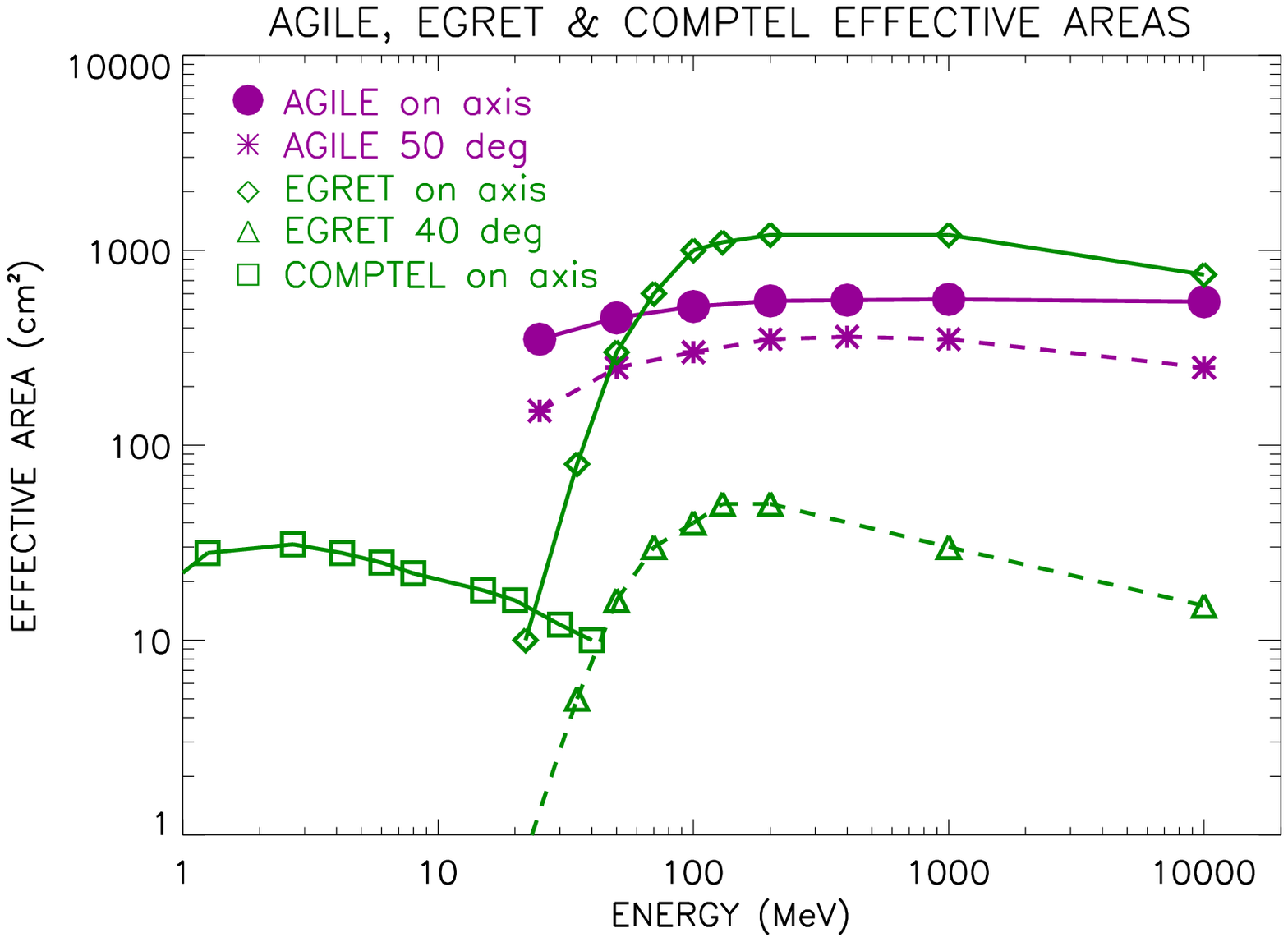}
\includegraphics[width=85mm,height=100mm]{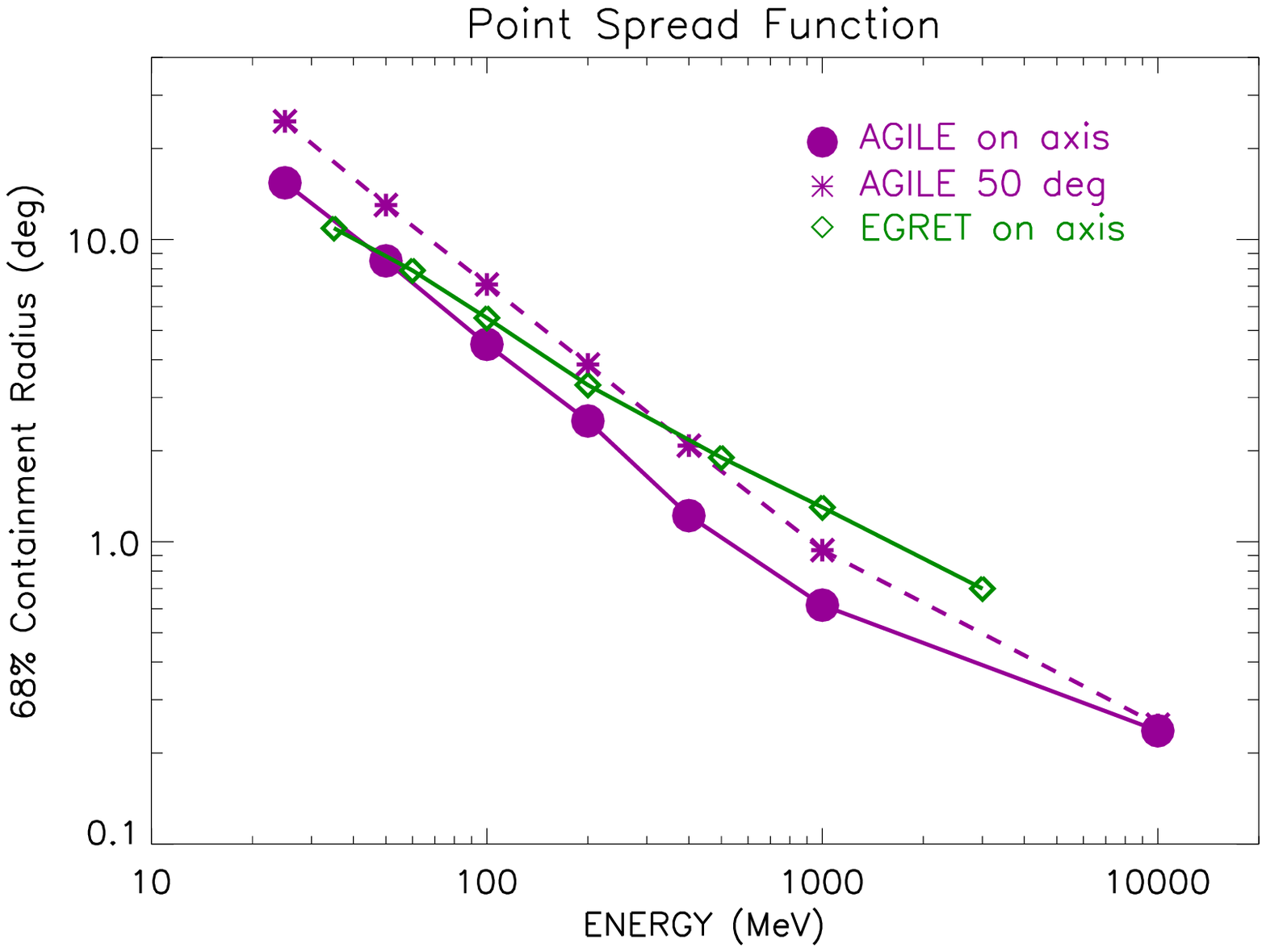}
\vspace*{-1.2cm}
\caption{~Effective area (left panel) and point spread function (right panel) as a function
of photon energy for AGILE and EGRET. The great advantage of AGILE vs. EGRET will be its very
large and almost ``flat-sensitivity'' field of view. AGILE will suffer only a moderate loss
of effective area and angular resolution at large incidence angles (up to $\sim$60$^{\circ}$).}
\end{figure}

\medskip

\section*{PULSAR STUDIES}
Among the $\sim$1500 rotation-powered pulsars, observed mainly in the radio band, only
7 are seen as pulsed gamma-ray sources, and another $\sim$10 have been reported as possible
detections (e.g. Kanbach, 2002). The availability of a wider 
and deeper sample of gamma-ray observations of pulsars could better constrain 
emission models that predict different beaming angles and directions for the 
radio and gamma-ray emission. 
AGILE will improve photon statistics for gamma-ray pulsation searches and will 
offer the first possibility, before the launch of GLAST (Michelson, 2001), of
detecting several young and energetic radio pulsars that have been discovered
since the end of the CGRO mission.

\begin{figure}
\hspace*{1.9cm}
\vspace*{-1.6cm}
\includegraphics[width=140mm,height=97mm]{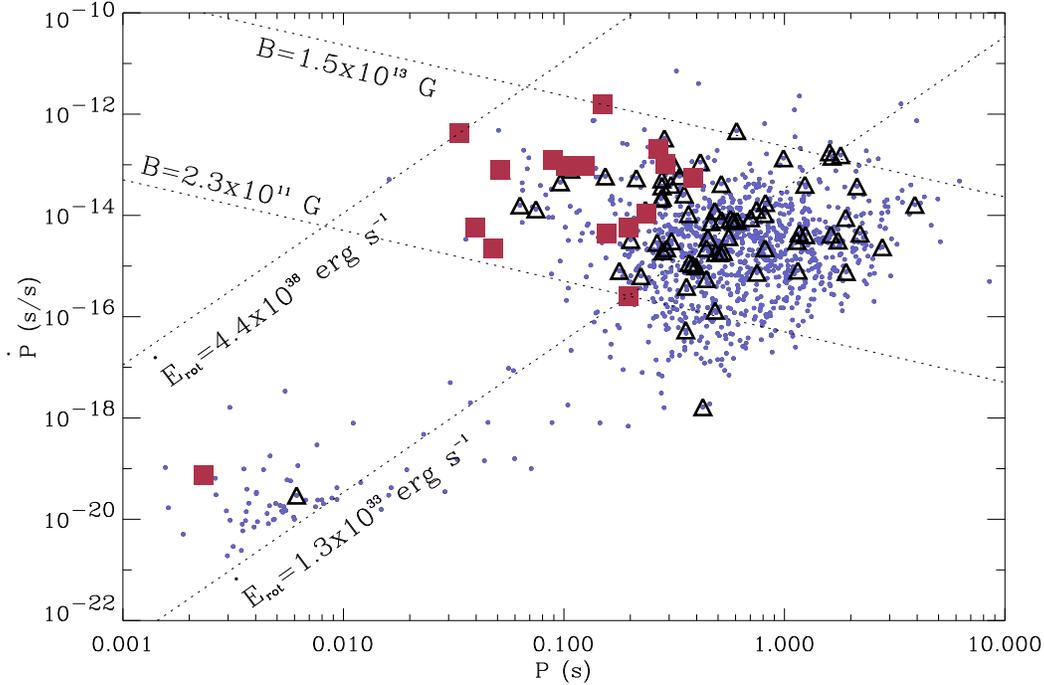}
\bigskip
\caption{~$P-\dot{P}$ diagram of a set of $\sim$1300 radio pulsars (dots) whose parameters are
obtained from public archives. Dotted lines correspond to the extreme values of $B$ and 
$\dot{E}_{rot}$ (excluding ms pulsars) of the $\sim$15 known and candidate
gamma-ray pulsars (filled squares). The ``gamma-ray pulsars region'' (upper rhomboidal region)
includes $\sim$400 radio pulsars likely to be efficient high-energy photons emitters. In this region
there are $\sim$20 non-variable EGRET unidentified sources (triangles) coincident with one or
more radio pulsars (the total number of radio/gamma coincidences being $\sim$40).}
\end{figure}

Except for the claimed detection of the millisecond pulsar J0218+4232 at high-energy 
gamma rays (Kuiper et al. 2000, 2002), all other
known gamma-ray pulsars discovered by EGRET (including low significance pulsed detections)
lie in the region of the $P-\dot{P}$ diagram
characterized by $B$$>$$2\times10^{11}$ $G$ and $\dot{E}_{rot}$$>$$1.3\times10^{33}$ $erg$ $s^{-1}$ (Figure 2). 
Radio pulsars in this region are natural gamma-ray pulsar candidates to be detected by instruments
like AGILE with sensitivity slighly better than EGRET. It is worth noting that tens of them
are within the error boxes of selected EGRET unidentified sources that do not show significant flux 
variations (compared to blazars) and are therefore good gamma-ray pulsar candidates.
Although many of these associations are expected to be chance coincidences, they represent
top-ranked targets for pulsations searches with AGILE.

The large field of view of AGILE allows simultaneous monitoring of many of these
sources, and could cover the whole sky with only 6 pointings (Figure 3). In
particular, a single pointing at the Galactic Center region could include tens of
nearby radio pulsars belonging to the ``gamma-ray pulsar region'' described in Figure 2,
 and $\sim$50 unidentified EGRET sources possibly
associated with radio-quiet pulsars, including those within the Gould Belt region
 (Gehrels et al. 2000).

AGILE will continuously observe pulsars for long periods ($\sim$1-2 months) avoiding
typical timing analysis problems arising from merging many short observations at 
different times. A typical exposure on the galactic plane after two years of observations
 is $\sim$2$\times$10$^{9}$ cm$^{2}$ s for $E$$>$100 MeV and the expected pulsar counts 
at these energies (e.g. $\sim$7000 cts from Geminga and $\sim$17000 from the Vela pulsar)
are comparable or slighly better than the corresponding values for 
EGRET.
The expected typical sensitivity after folding data at known pulsar periods is 
$\sim$5$\times$10$^{-8}$ ph/cm$^{2}s$ for $E$$>$100 MeV.
Furthermore, using accurate radio ephemerides, it is possible to extract pulsar
photons also in confused regions and exploit low energy data (20-100 MeV) for which
angular resolution is poor but the AGILE effective area is still useful and better
than that of EGRET (Figure 1).

In contrast to the relation valid at soft X-ray energies ($ F_{x}\propto \dot{E} d^{-2}$ ),
the expected gamma-ray flux of radio pulsars is directly correlated to the Goldreich-Julian
current and can be estimated according to the law
$ F_{\gamma}\propto \sqrt{\dot{E}} d^{-2}$ (Kanbach, 2002) fitting EGRET pulsars well.
The dispersion of the fit provides the min/max normalization values allowing a worst/best
case approach for the gamma-ray fluxes estimates ($F_{min}$ and $F_{max}$ in Table 1)
in the assumption that the gamma-ray beams intersect the line of sight.
In this way it is possible to extract a subsample of radio-loud pulsars 
expected to be above the AGILE sensitivity threshold. Up to $\sim$100 radio pulsars (Figure 3) could
provide enough photons to be
in principle observable by AGILE, the actual number of detections being dependent on the emission geometry.
Assuming, for example, gamma-ray emission cones of $\sim$1 sr with random orientation respect to the 
radio beam, roughly $\sim$10 sources of our sample should be detectable with AGILE.
Table 1 provides the first 35 top ranked radio-pulsars likely to have detectable fluxes 
($F_{min}$$>$$2\times10^{-8}$ $ph/cm^{2} s$) even in the absence
of coincidences with EGRET sources. It also includes (lower part)
all the other weaker
11 radio-pulsars of the ``gamma-ray pulsar region'' coincident with EGRET unidentified sources and with
expected fluxes near AGILE sensitivity.

Apart from the sources within the ``gamma-ray pulsar region'', AGILE will have the first opportunity to 
explore the ``millisecond region'' of the $P-\dot{P}$ diagram
 looking for peculiar gamma-ray emission from these objects and to confirm the
claimed detection in high-energy gamma rays of PSR J0218+4232 (Kuiper et al. 2000, 2002).

More detailed simulations of the population of gamma-ray pulsars, including deep
statistical analysis of the emission geometry and beaming, predict very different
numbers of radio-loud and radio-quiet pulsars detectable in gamma rays, strongly depending
 on the assumed model
(see e.g. Gonthier et al. 2001; Mac Laughlin and Cordes, 2001). 
Extrapolating these results to the AGILE wide-band sensitivity, we expect to discover 
at least a dozen new objects.


\begin{figure}
\hspace*{2.1cm}
\vspace*{-1.cm}
\includegraphics[width=133mm,height=85mm]{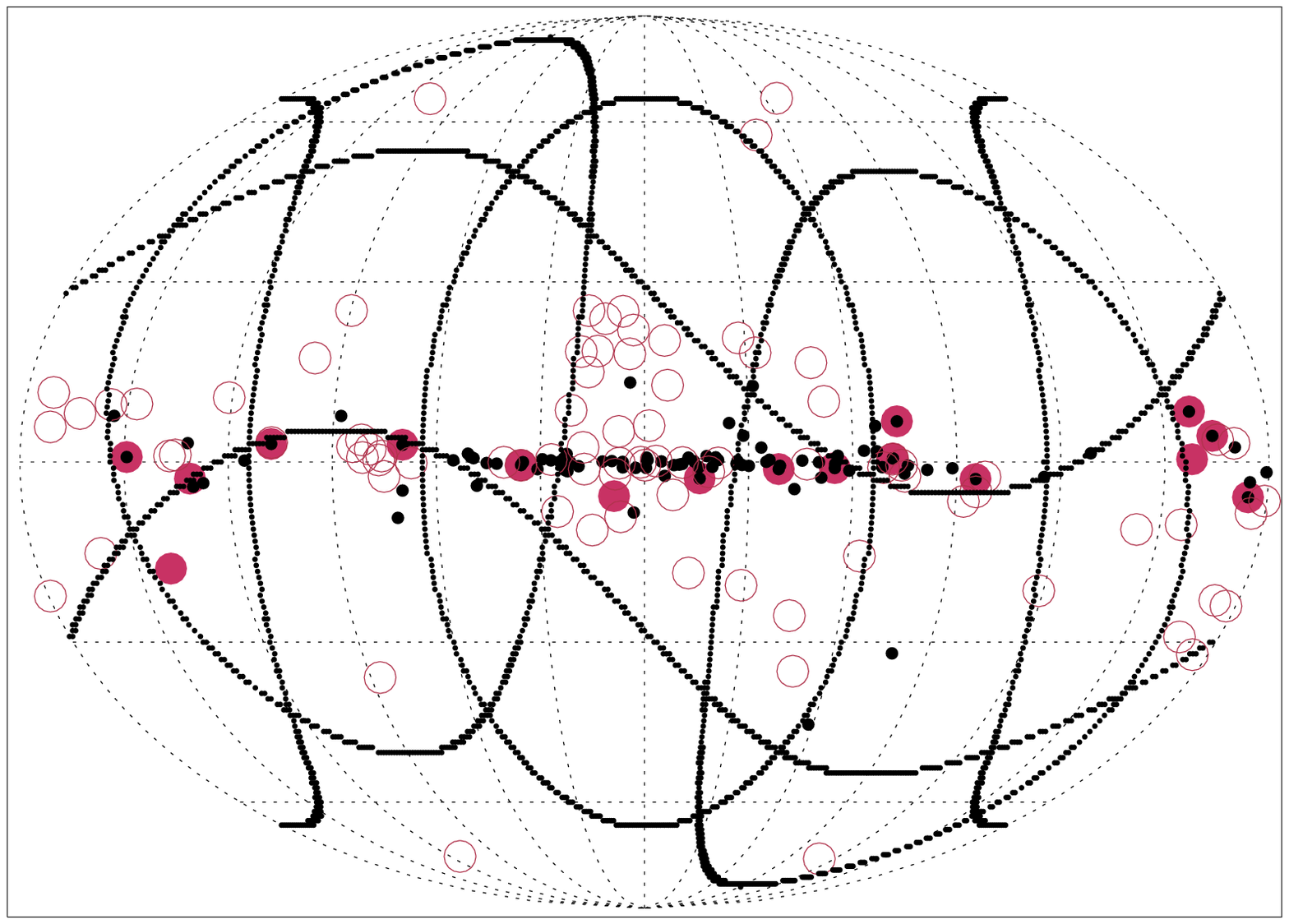}
\caption{~Map in Galactic coordinates (Galactic Centre in the 
centre of the figure) of the $\sim$100 known radio pulsars (dots) belonging to
the ``gamma-ray pulsar region'' described in Figure 2 and with maximum expected gamma-ray flux near
AGILE sensitivity (see text).
The map also shows claimed gamma-ray pulsars 
(filled circles) and non-variable unidentified EGRET sources (circles) likely 
to consist in part of radio-quiet pulsars. Bold contours indicate the fields of view of
a set of 6 AGILE pointings covering the whole sky.}
\end{figure}

\begin{table*}
\vspace*{-1.0cm}
\begin{center}
\caption[]{~Preliminary sample of the ranking list of radio-pulsars expected to have gamma-ray fluxes near
 AGILE sensitivity (see text). This sample includes 14 radio-pulsars coincident with selected unidentified
EGRET sources, some of which have marginal pulsed detections.
$F_{min}$ and $F_{max}$ represent the expected gamma-ray flux ranges of the sources
in the hypothesis that the gamma-ray beams intersect the line of sight.}
\small
\begin{tabular}{cccccccccccc}
\hline
\noalign {\smallskip}
PSR Name & Gal. Long. & Gal. Lat. & $P$ & $\dot{P}$ & $D$ & $\dot{E}$ & $F$$_{\rm{min}}$$^{*}$  & $F$$_{\rm{max}}$$^{*}$ & 3EG coinc. \\
&           {\footnotesize deg}    &  {\footnotesize deg}   &  {\footnotesize s} & {\footnotesize 10$^{-15}$ s/s} & {\footnotesize kpc} & {\footnotesize log erg/s} \\
\noalign {\smallskip}
\hline
\noalign {\smallskip}

   J1932+1059 &        47.38&        -3.88&        0.227&           1.16&         0.17&        33.59&          109&          986&                               \\
   J0659+1414$^{\#}$ &       201.11&         8.26&        0.385&          55.01&         0.76&        34.58&           16&          151&                               \\
   J2043+2740 &        70.61&        -9.15&        0.096&           1.26&         1.13&        34.75&            9&           83&                               \\
   J1908+0734 &        41.58&        -0.27&        0.212&           0.82&         0.58&        33.53&            8&           76&                               \\
   J1048-5832$^{\#}$ &       287.42&         0.58&        0.124&          96.32&         2.98&        36.30&            8&           71&                3EG J1048-5840 \\
   J1524-5625 &       323.00&         0.35&        0.078&          38.95&         3.84&        36.51&            6&           54&                               \\
   J0742-2822 &       243.77&        -2.44&        0.167&          16.81&         1.89&        35.16&            5&           47&                               \\
   J0117+5914 &       126.28&        -3.46&        0.101&           5.85&         2.14&        35.34&            5&           46&                               \\
   J1826-1334 &        18.00&        -0.69&        0.101&          75.49&         4.12&        36.46&            4&           44&                               \\
   J1531-5610 &       323.90&         0.03&        0.084&          13.74&         3.10&        35.96&            4&           44&                               \\
   J1809-1917 &        11.09&         0.08&        0.083&          25.54&         3.71&        36.25&            4&           43&                               \\
   J1617-5055 &       332.50&        -0.27&        0.069&         137.20&         6.46&        37.21&            4&           43&                               \\
   J1803-2137 &         8.40&         0.15&        0.134&         134.43&         3.94&        36.35&            4&           43&                               \\
   J1918+1541 &        49.89&         1.36&        0.371&           2.54&         0.68&        33.29&            4&           42&                               \\
   J0205+6449 &       130.72&         3.08&        0.066&         193.93&         7.54&        37.43&            4&           41&                               \\
   J1856+0113$^{\#}$ &        34.56&        -0.50&        0.267&         208.41&         2.78&        35.63&            4&           38&                3EG J1856+0114 \\
   J1913+1011 &        44.48&        -0.17&        0.036&           3.37&         4.48&        36.46&            4&           37&                               \\
   J0454+5543 &       152.62&         7.55&        0.341&           2.37&         0.79&        33.37&            3&           34&                               \\
   J0940-5428 &       277.51&        -1.29&        0.088&          32.87&         4.27&        36.29&            3&           34&                               \\
   J1801-2451 &         5.25&        -0.88&        0.125&         127.98&         4.61&        36.41&            3&           34&                               \\
   J0538+2817 &       179.72&        -1.69&        0.143&           3.66&         1.77&        34.69&            3&           31&                               \\
   J1825-0935 &        21.45&         1.32&        0.769&          52.36&         1.01&        33.66&            3&           29&                               \\
   J1549-4848 &       330.49&         4.30&        0.288&          14.11&         1.54&        34.37&            3&           28&                               \\
   J1718-3825 &       348.95&        -0.43&        0.075&          13.22&         4.24&        36.10&            3&           27&                3EG J1714-3857 \\
   J1730-3350 &       354.13&         0.09&        0.139&          85.10&         4.24&        36.09&            3&           27&                               \\
   J1705-1906 &         3.19&        13.03&        0.299&           4.14&         1.18&        33.79&            2&           25&                               \\
   J1420-6048 &       313.54&         0.23&        0.068&          83.17&         7.69&        37.02&            2&           24&                               \\
   J0358+5413$^{\#}$ &       148.19&         0.81&        0.156&           4.40&         2.07&        34.66&            2&           22&                               \\
   J1509-5850 &       319.97&        -0.62&        0.089&           9.17&         3.81&        35.71&            2&           22&                               \\
   J1739-3023 &       358.09&         0.34&        0.114&          11.40&         3.41&        35.48&            2&           21&                               \\
   J1835-1106 &        21.22&        -1.51&        0.166&          20.61&         3.09&        35.25&            2&           19&                               \\
   J1530-5327 &       325.33&         2.35&        0.279&           4.68&         1.46&        33.93&            2&           19&                               \\
  J1302-6350$^{\#}$ &       304.18&        -0.99&        0.048&           2.28&         4.60&        35.92&            2&           19&                               \\
   J2337+6151 &       114.28&         0.23&        0.495&         191.81&         2.47&        34.79&            2&           18&                               \\
   J1453-6413 &       315.73&        -4.43&        0.179&           2.75&         1.84&        34.27&            2&           18&                               \\
\noalign {\smallskip}
\hline
\noalign {\smallskip}
   J1837-0604 &        25.96&         0.26&        0.096&          45.17&         6.19&        36.30&            1&           16&                3EG J1837-0606 \\
   J2229+6114$^{\#}$ &       106.65&         2.95&        0.052&          78.27&        12.00&        37.35&            1&           14&                3EG J2227+6122 \\
   J1105-6107 &       290.49&        -0.85&        0.063&          15.83&         7.07&        36.39&            1&           14&                3EG J1102-6103 \\
   J1740-3015 &       358.29&         0.24&        0.607&         465.87&         3.28&        34.92&            1&           11&                3EG J1744-3011 \\
   J1637-4642 &       337.79&         0.31&        0.154&          59.20&         5.77&        35.81&            1&           10&                3EG J1639-4702 \\
   J1745-3040 &       358.55&        -0.96&        0.367&          10.66&         2.08&        33.93&            1&            9&                3EG J1744-3011 \\
   J1016-5857 &       284.08&        -1.88&        0.107&          80.62&         9.31&        36.41&            0&            8&                3EG J1013-5915 \\
   J1757-2421 &         5.31&         0.02&        0.234&          13.00&         3.50&        34.60&            0&            7&                3EG J1800-2338 \\
   J1853+0056 &        34.02&        -0.04&        0.276&          21.39&         3.82&        34.61&            0&            6&                3EG J1856+0114 \\
   J1715-3903 &       348.10&        -0.32&        0.278&          37.69&         4.80&        34.84&            0&            5&                3EG J1714-3857 \\
   J0614+2229 &       188.79&         2.39&        0.335&          59.63&         4.72&        34.80&            0&            5&                3EG J0617+2238 \\
  
\noalign {\smallskip}
\hline
\label{tab:spec}
\end{tabular}
\normalsize
\end{center}
\vspace{-0.5 cm}
{\small *) 10$^{-8}$ ph/cm$^{2}$ s ($E$$>$100 MeV)} \\
{\small $\#$) Low significance pulsed detection with EGRET} 
\end{table*}


\section*{DISCUSSION}
AGILE will contribute to the study of known and newly discovered gamma-ray pulsars in
several ways:
1) detecting possible secular variations of the gamma-ray emission; 2) studying unpulsed
gamma-ray emission from plerions in supernova remnants and searching for time variability
of pulsar/wind nebula interactions (e.g. in Crab); 3) looking for possible breaks in pulsar
spectra; 4) obtaining high time resolution of known gamma-ray pulsars.
In particular, AGILE will have the first opportunity to observe pulsars with better time 
resolution than EGRET, looking for possible microstructures in folded gamma-ray light
curves such as those clearly seen in single radio pulses from Vela (Johnston et al. 2001).
We simulated a $\sim5$$\times$$10^{6}$ $s$ AGILE observation of the Vela pulsar including different microstructure templates
in the light curve.
A preliminary analysis
shows that we can achieve an effective resolution better than $\sim$50 $\mu$s with a good signal-to-noise
ratio allowing us to discover hypothetical microstructures not resolved with lower resolution (Figure 4).
Compared to the pulsar population as a whole, young and energetic 
radio-loud pulsars, of which the gamma-ray pulsars are
a subset, present a significant amount of timing noise and occasional glitches. 
Since these strongly affect the folding process for long exposures and
smear gamma-ray light curves, simultaneous radio observations are essential
for accurate gamma-ray timing analysis of these sources.


The analysis of a larger sample of pulsar spectral breaks could provide a 
powerful tool to constrain the production processes and sites in the magnetosphere (e.g. the height of the
 emission region 
from the neutron star surface). Pulsar spectra are generally curved in the gamma-ray band leading to a
cut-off at a few GeV and therefore very difficult to investigate with
reasonable energy resolution ($\Delta$$E/E$$\sim1$) due to the poor photon statistics
 available at these
energies. 
However, PSR B1509-58, a pulsar with a very strong magnetic field ($>$$10^{13}$ $G$), shows a spectral
turnover in the 10 MeV range (Kuiper et al. 1999) which can be explained with photon
absorption by photon splitting in the scenario of polar cap models (Harding et al. 1997). For the competing
 scenario with outer gaps, Zhang and Cheng (2000)  offer an alternative explanation.
There are now more than $\sim$30 pulsars known with $B>10^{13} G$. These highly-magnetized pulsars
represent a good opportunity for AGILE to perform a set of gamma-ray observations providing spectral 
break measurements in a more accessible energy range, possibly providing conclusive evidence for the 
signature of this exotic photon splitting process in the strongest magnetic fields of pulsars.



\begin{figure}
\vspace*{-3.0cm}
\includegraphics[width=150mm,height=204mm,angle=-90]{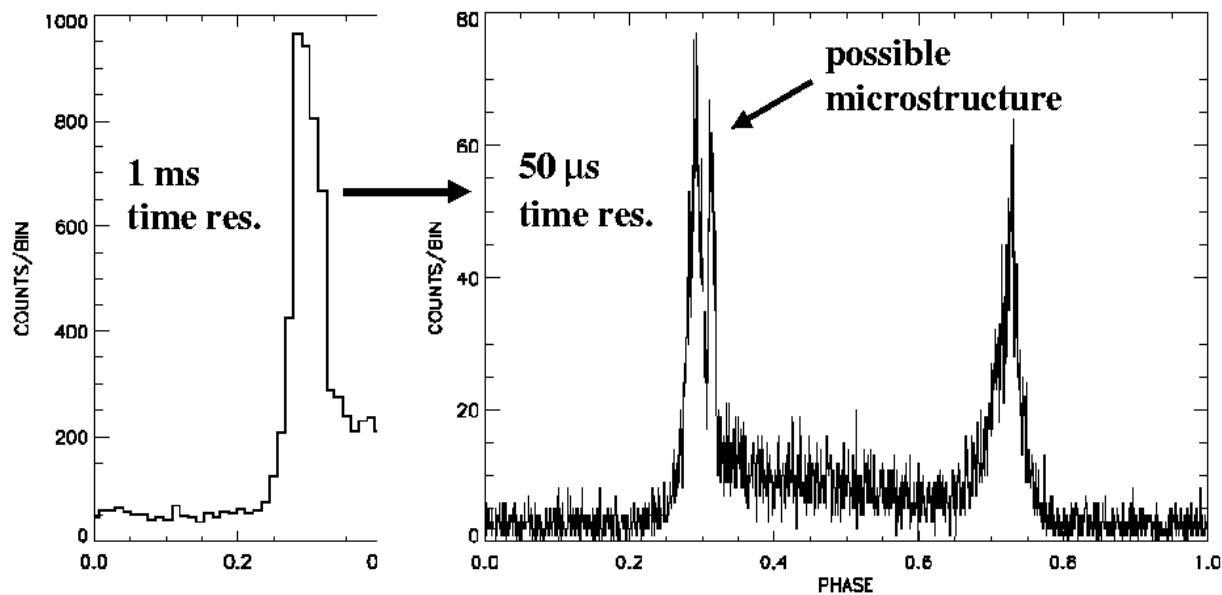}
\vspace*{-4cm}
\caption{~Simulation of an AGILE observation of the Vela pulsar. Timing analysis with resolution 
10-50 $\mu$s could reveal possible microstructures within the peaks of the light curve.}
\end{figure}

\bigskip
\medskip
\noindent
E-mail address of A. Pellizzoni~~~alberto@mi.iasf.cnr.it

\end{document}